\documentclass[psfig]{elsart}
\IfFileExists{srcltx.sty}{\usepackage[active]{srcltx}}%

\usepackage{graphicx}% Include figure files
\usepackage{dcolumn}% Align table columns on decimal point
\usepackage{epsf}
\usepackage{bm}% bold math
\usepackage{amsmath}
\usepackage{rotate}
\usepackage{rotating}

\newcommand{\beq}{\begin{equation}}
\newcommand{\eeq}{\end{equation}}
\newcommand{\bea}{\begin{eqnarray}}
\newcommand{\eea}{\end{eqnarray}}

\begin{document}
\begin{frontmatter}

%\vskip 1.0truecm
%\begin{flushright}
%{UCLA/05/TEP/23}
%\end{flushright}

\title{A new bound on supersymmetric Q-balls} 

\author{Jason Schissel} 

%\affiliation{
\address
%\thanks
{Department of Physics and Astronomy,
UCLA, Los Angeles, CA 90095-1547, USA } 
%\vspace{0.5truecm}

\begin{abstract}

Stable baryonic Q-balls, which appear in supersymmetric extensions of the
Standard Model, could form at the end of cosmological inflation from
fragmentation of the Affleck-Dine condensate.  They can be dark matter. The
existing bounds rely on the ability of Super-Kamiokande detector to trigger on
a slowly moving bright source, which may be difficult for techinical reasons. 
We present a weaker but more robust bound based on the flux of neutrinos
produced by relic Q-balls interacting in Earth. 

\end{abstract}

\end{frontmatter}

In the search for dark matter, supersymmetry (SUSY) provides two possible
candidates: the lightest supersymmetric particle (for example, neutralino or
gravitino), and stable Q-balls. SUSY Q-balls are nontopological solitons that
carry baryon number~\cite{Kusenko:1997}. Some Q-balls are stable or have
lifetimes in excess of the age of the universe in theories with gauage-mediated
SUSY breaking.   They can form in the early universe
from the fragmentation of the Affleck--Dine condensate~\cite{Affleck:1984fy}
and they can presently exist as dark 
matter~\cite{Kusenko:1997si,Dine:2003ax}.  

Because of the wide acceptance of supersymmetry as a candidate for
physics beyond the standard model-- and the corresponding implication of the
existence of Q-balls-- the desire to find experimental evidence for Q-balls is
strong. However, there is a large range of uncertainty in the parameters
involved , such as Q-ball mass, radius, cosmic abundance, and cross section. We
therefore attempt to use extant experimental data to bound the region of future
experimental interest.

Interactions of Q-balls in matter result in induced "proton decay" processes. 
A quark scattering off a Q-ball can reflect back as an antiquark with
probability of order one~\cite{Kusenko:2004yw}.  The baryon number of the
Q-ball changes in the process, so the overall baryon number of Q-ball and
 hadron is unchanged.  However, a proton scattering off Q-ball can turn
into an antiproton and can annihilate with one of the baryons in the
ambient matter.  Hence, a SUSY Q-ball passing through matter deposits as much
as $\sim 10^3$~GeV of energy per 1~cm of track, and most of  this energy is
generated in the form of soft pions \cite{bh}.

Astrophysical bounds based on stability of neutron stars are the strongest
limits on many types of Q-balls~\cite{kls}.  A neutron star can be eaten away
in a time shorter than the age of the universe if the Q-balls can grow big in
its interior. However, depending on the structure of the operators that lift
the MSSM flat directions, the baryon number violation can be triggered by a
large VEV, which can stymie the further growth of Q-balls.  For such Q-balls,
the laboratory bounds are still the strongest.  

The current laboratory bounds~\cite{kkst,Arafune:2000yv} rely on the
abundance of relic SUSY Q-balls rely on the ability of Super-Kemiokande
detector to register the passage of a slowly moving object producing a large
amount of light.  On enetering the detector, the Q-ball would cause all its
phototubes to saturate.  Since all the tubes are effectively blinded, no
further information is available.  Events of this kind have happened in the
past, both in Super-Kamiokande and in its predecessor, 
Kamiokande~\cite{Hirata:1992sc}. A clustering of low energy spallation events
followed a bright flash saturating the detector's phototubes. A Q-ball
interacting within the detector could produce such results. To establish that
this is not a SUSY Q-ball, one would have to analyze the data during the 0.1
millisecond time window subsequent to the onset of the bright flash. Relic
Q-balls are expected to have a velocity $v\sim 10^{-3}c$. However,
Super-Kamiokande routinely discards the data in the wake of various events
because the signal would be plagued by multiple reflections of light inside the
detector. Therefore, there is a possibility that the passages of SUSY Q-balls
are not registered by Super-Kamiokande, except as events of the kind described
in Ref.~\cite{Hirata:1992sc}.  Similar effects have been reported by
Pamir~\cite{pamir} and other experiments~\cite{centauro}. 

It is useful, therefore, to establish an independent limit even if such a
limit is weaker than those presented in Refs.~\cite{kkst,Arafune:2000yv}.  We
will set a new limit based on the neutrinos SUSY Q-balls would produce
interacting inside the Earth. 

Several assumptions are made about Q-ball characteristics. A good review of the mathematics behind these assumptions can be found at Ref. \cite{ad}.

1) In order to be a dark matter candidate, Q-balls must have baryon number Q$\sim 10^{24\pm5}$

2) M(Q) $\sim \rm{M_{SUSY}} \rm{Q}^{3/4}$ 

3)$\rm{M_{SUSY}} \sim 1-10 \rm{TeV}$

4) $\frac{\rm{M(Q)}}{\rm{Q}} < \rm{M_{proton}}$  

Condition (4) is a stability condition; if Q-ball mass per baryon number is greater than the proton mass, the Q-ball is not stable and may decay into proton(s).

Taken together, these assumptions imply a Q-ball mass less than $10^{20}$ eV. Such Q-balls could exist in the form of relics from cosmological inflation. It is also possible that smaller Q-balls could be astrophysically accelerated to comparable energies.

The manner of Q-ball interaction with matter as described by Kusenko, et. al.
is that Q-balls reflect incoming quarks as antiquarks with probability almost
one \cite{Kusenko:2004yw}.  
In the Earth, a reflected antiproton will interact almost immediately with
the surrounding matter. Proton-antiproton interactions in this energy regime
are fairly well understood. At these energies, the interactions of neutrons are not significantly different from those of protons\cite{hillaspaper}. Therefore the approximation of a protonic earth is good.

Proton-antiproton interactions will create a host of pions. These pions will have time to interact with the surrounding matter before they decay, but at these (low) energies, only hard scatterings will occur. These scatterings will cause the charged pions to radiate some electromagnetic energy, but ultimately they will still decay into neutrinos. The neutrinos will escape from within the Earth and pass through surface neutrino detectors. 
Other products of p-pbar annihilation, such as photons and electrons, will be stopped by the Earth.

By comparing expected total neutrino flux to detector sensitivity, neutrino nondetection will set an upper bound on Qball abundance. Specifically, we know the detected flux of atmospheric neutrinos at the Super-K detector, which will act as a background to the desired detection, thus also as a benchmark for our predicted flux.

We will proceed through dimensional argument.

Flux is defined as the number of particles traversing a square area per unit
time. We can find the flux of atmospheric neutrinos through Super-K at Ref. \cite{hkkm}. At 200 MeV, it is about $6 \rm{cm^{-2}s^{-1}}$. 

In the steady state, the rate of neutrino production within the volume of the
Earth ($\rm{r}_{\nu}$) must equal neutrino flux at the Earth's surface.

\beq
4\pi \rm{R_{Earth}}^2 \cdot \rm{F}_{\nu} = r_{\nu}
\eeq

The production rate of neutrinos within the Earth depends on the unknown Q-ball flux, $\rm{F_{Q}}$.

We can convert the unknown Q-ball flux at the detector into a Q-ball density in space by dividing by the velocity of the Q-ball, $\rm{v_{Q}}$.

Given a density, we can multiply by the volume of the Earth, $\rm{V_{E}}$ to get the total number of Q-balls in the Earth at a given time.

Multiplying the total number of Q-balls by the number of neutrinos per interaction times interactions per second gives the total number of neutrinos per second in the volume.

In the approximately uniform matter of the Earth, each second there are 

\beq
\sigma \cdot \rm{n_{p}} \cdot \rm{v}
\eeq

Q-ball interactions per second, where $\rm{n_{p}}$ is the proton number density, v is the velocity of the Q-ball, and the cross section is geometric

\beq
\sigma = \pi \rm{R_{Q-ball}}^{2}
\eeq

The radius of the Q-ball
\beq
\rm{R_{Q-ball}} = \rm{\frac{Q^{1/4}}{M_{SUSY}}} = 10^{-6} \rm{eV}^{-1}
\eeq

In the Earth, where $\rm{n_{p}} \sim 3\cdot 10^{24} \rm{cm}^{-3}$, the number of interactions per second is then

\beq
 \pi \cdot(10^{-6} eV^{-1})^{2} \cdot  3\cdot 10^{24} \rm{cm}^{-3} \cdot \rm{v} \rm{\frac{cm}{s}} 
\eeq

In each interaction

\beq
p + \overline{p} \rightarrow 5\cdot(\pi_{+} +\pi_{-} + \pi_{0})\cdot \frac{1}{3}
\eeq

5 pions are produced on average \cite{bh}.

The pions share the energy of the 2 GeV annihilation (of the proton and anti-proton nearly at rest), resulting in about 400MeV/pion. As
the charged pions decay (primary decay mode, 99.99\%), neutrinos are produced
in two subsequent steps

\beq
 \pi_{+} \rightarrow \mu_{+} \nu_{\mu}
\eeq
\beq
 \mu_{+} \rightarrow e_{+} \overline{\nu_{e}} \nu_{\mu}
\eeq

Resulting in an average of 3.33 muon neutrinos at 200MeV and additional mu- and nu- neutrinos at about 1\
/3 this energy.

Finally, dividing by the surface area of the Earth, $\rm{A_{E}}$, we have the appropriate dimensions of $\rm{m^{-2}s^{-1}}$. 

Algebraically, 
\beq
\frac{\rm{F_{Q}} \cdot \rm{V_{E}} \cdot \sigma \cdot \rm{n_{p}} \cdot \rm{v} \cdot 3.33}{\rm{v}\cdot\rm{A_{E}}}=\rm{r_{\nu}}
\eeq

Solving for $\rm{F_{Q}} $
\beq
\rm{F_{Q}} = \frac{3}{6.4\cdot 10^{8}\rm{cm}}\cdot     \frac{6 \rm{cm^{-2}s^{-1}}}{3.33\cdot 10^{-30}\rm{cm^{2}}\cdot10^{24}\rm{cm^{-3}}}
\eeq

where we assume a spherical earth for purposes of volume and surface area calculation. This calculation is \emph{independent of the velocity of the Q-ball}, since that quantity cancels in the division.

Then we simply solve for $\rm{F_{Q}}$.
\beq
\rm{F_{Q}} \sim 10^{-2}\rm{cm^{-2}s^{-1}}
\eeq

Note that this argument remains good even in the relativistic case. For a gamma factor greater than 1, the output of such interactions will be beamed in a cone with opening angle $\sim \frac{1}{\gamma}$, but because the arrival direction of the Q-balls is presumed to be symmetric, equal numbers of neutrinos are gained and lost through this effect. 

 At the beginning of this paper, several assumptions about the nature of Q-balls were made. This is, therefore, a bound on the abundance of Q-balls which are dark matter candidates by virtue of their mass, radius, and baryon number, around a total energy of $10^{20}$eV. More energetic Q-balls, or less massive Q-balls with a higher gamma factor would produce more pions in their interactions with matter, resulting in more neutrinos and therefore a smaller upper bound.

\section{Conclusions}

Q-balls are a fundamental prediction of supersymmetry and have the power to explain many otherwise puzzling phenomena detected in the highest energy regimes. Certain types of Q-balls are good candidates for dark matter, but have a large range of possible physical parameters. We have set a bound on the flux of Q-balls through detectors that is independent of the detector ability to recognize Q-ball events.

The author thanks A.~Kusenko for suggesting the idea for this new limit, as 
well as for helpful discussions. This work was supported in part by the
U.S. Department of Energy grant DE-FG03-91ER40662 and by the NASA ATP grants
NAG~5-10842 and NAG~5-13399.


\begin{thebibliography}{99}
%1
\bibitem{Kusenko:1997}
  A.~Kusenko,
  %``Solitons in the supersymmetric extensions of the standard model,''
  Phys.\ Lett.\ B {\bf 405}, 108 (1997); 
  %[arXiv:hep-ph/9704273].
  %%CITATION = HEP-PH 9704273;%%
  %A.~Kusenko,
  %``Small Q balls,''
  Phys.\ Lett.\ B {\bf 404}, 285 (1997); 
  %[arXiv:hep-th/9704073].
  %%CITATION = HEP-TH 9704073;%%
G.~R.~Dvali, A.~Kusenko and M.~E.~Shaposhnikov,
  %``New physics in a nutshell, or Q-ball as a power plant,''
  Phys.\ Lett.\ B {\bf 417}, 99 (1998). 
%  [arXiv:hep-ph/9707423].
  %%CITATION = HEP-PH 9707423;%%

%2
\bibitem{Kusenko:1997si}
  A.~Kusenko and M.~E.~Shaposhnikov,
  %``Supersymmetric Q-balls as dark matter,''
  Phys.\ Lett.\ B {\bf 418}, 46 (1998).
  %[arXiv:hep-ph/9709492].
  %%CITATION = HEP-PH 9709492;%%

%3  
  \bibitem{Affleck:1984fy}
  I.~Affleck and M.~Dine,
  %``A New Mechanism For Baryogenesis,''
  Nucl.\ Phys.\ B {\bf 249}, 361 (1985).
  %%CITATION = NUPHA,B249,361;%%
 
%4 
\bibitem{Dine:2003ax} For review, see, e.g., 
 K.~Enqvist and A.~Mazumdar,
  %``Cosmological consequences of MSSM flat directions,''
  Phys.\ Rept.\  {\bf 380}, 99 (2003); 
  %[arXiv:hep-ph/0209244].
  %%CITATION = HEP-PH 0209244;%%
  M.~Dine and A.~Kusenko,
  %``The origin of the matter-antimatter asymmetry,''
  Rev.\ Mod.\ Phys.\  {\bf 76}, 1 (2004).
%  [arXiv:hep-ph/0303065].
  %%CITATION = HEP-PH 0303065;%%

%5
\bibitem{Kusenko:2004yw}
  A.~Kusenko, L.~Loveridge and M.~Shaposhnikov,
  %``Supersymmetric dark matter Q-balls and their interactions in matter,''
  Phys.\ Rev.\ D {\bf 72}, 025015 (2005)
  [arXiv:hep-ph/0405044].
  %%CITATION = HEP-PH 0405044;%%
%6  
\bibitem{bh} W. Blumel, U. Heinz. Z.Phys. C67 (1995) 281-286. %Pion Multiplicity Distriubtion in Proton-Antiproton Annihilation at Rest

%7
\bibitem{kls}
 A.~Kusenko, L.~C.~Loveridge and M.~Shaposhnikov,
  %``Astrophysical bounds on supersymmetric dark-matter Q-balls,''
  JCAP {\bf 0508}, 011 (2005)
  [arXiv:astro-ph/0507225].
  %%CITATION = ASTRO-PH 0507225;%%

%8
\bibitem{kkst} A. Kusenko, V. Kuzmin, M. Shaposhnikov, P.G. Tinyakov. 
Phys. \ Rev. \ Lett. {\bf 80} (1998)

%9
\bibitem{Hirata:1992sc}
  K.~S.~Hirata {\it et al.},
  % ``Search For Neutrino Induced Low-Energy Electron Event Clusters In
  % Kamiokande-II,''
  Phys.\ Rev.\ D {\bf 45}, 3355 (1992).
  %%CITATION = PHRVA,D45,3355;%%
%10
\bibitem{Arafune:2000yv}
  J.~Arafune, T.~Yoshida, S.~Nakamura and K.~Ogure,
  %``Experimental bounds on masses and fluxes of nontopological solitons,''
  Phys.\ Rev.\ D {\bf 62}, 105013 (2000)
  [arXiv:hep-ph/0005103].
  %%CITATION = HEP-PH 0005103;%%
%11
\bibitem{pamir} S.~G.~Bayburina {\it et al}., Proc. 16th Intern. Cosmic Ray
  Conf., {\bf 7}, 279 (1979).

  
\bibitem{centauro} C.~M.~G.~Lattes, Y.~Fujimoto, and S.~Hasegawa,
Phys. Rep. {\bf 65} 151 (1980).


\bibitem{ad} I. Affleck, M. Dine, Nucl. Phys. B 249, 361 (1985); M. Dine, L. Randall, S. Thomas, Phys.Rev. Lett. 75, 398 (1995); Nucl. Phys. B 458, 291 (1996); R. Allahverdi, B. A. Campbell and J. R. Ellis, Nucl. Phys. B 579, 355 (2000); A. Anisimov and M. Dine, Nucl. Phys.B 619, 729 (2001); A. Anisimov, Phys. Atom. Nucl. 67, 640 (2004) [Yad. Fiz. 67, 660 (2004)]; M. Kawasaki, K. Konya and F. Takahashi, Phys. Lett. B 619, 233 (2005).
  
\bibitem{hillaspaper} P. Bhattacharjee and G. Sigl. Phys.Rep. (1999)
[arXiv:astro-ph/9811011].

\bibitem{hkkm}Honda, M, T Kajita, K Kasahara, and S Midorikawa. Calculation of the Flux of Atmospheric Neutrinos. Phys. Rev (1995). http://xxx.lanl.gov/hep-ph/9503439.

\bibitem{NA} Na49 Collaboration. Energy Dependence of Pion and Kaon Production in Central Pb+Pb Collisions. (2002). 
http://arxiv.org/nucl-ex/0205002.


\bibitem{dk} M. Dine and A. Kusenko, Rev. Mod. Phys. 76, 1 (2004).

\bibitem{em}K. Enqvist and A. Mazumdar, Phys. Rept. 380, 99 (2003).

\bibitem{det} J. Ahrens, et. al
May 2003
arXiv:astro-ph/0305196 v1 


\bibitem{kls2} A. Kusenko, L. Loveridge, M. Shaposhnikov, arXiv:hep-ph/0405044

\bibitem{gzks} K. Greisen, Phys. Rev. Letters 16 (1966) 748;
G.T. Zatsepin and V.A. Kuz'min, JETP Letters 4 (1966) 78.
F.W. Stecker, Phys. Rev. Letters 21 (1968) 1016.

\bibitem{particles} V.~Berezinsky, M.~Kachelriess, and A.~Vilenkin,
Phys. Rev. Lett. {\bf 79}, 4302 (1997); V.~A.~Kuzmin and V.~A.~Rubakov,
Phys. Atom. Nucl. {\bf 61}, 1028 (1998) [Yad. Fiz. {\bf 61}, 1122 (1998)];
M.~Birkel and S.~Sarkar, Astropart. Phys. {\bf 9}, 297 (1998); K.~Benakli,
J.~Ellis, and D.~V.~Nanopoulos, Phys. Rev. {\bf D59}, 047301 (1999);  V.~Kuzmin
and I.~Tkachev, JETP Lett. {\bf 68}, 271 (1998); Phys. Rev. {\bf D59}, 123006
(1999); G.~Gelmini and A.~Kusenko,
  %``Unstable superheavy relic particles as a source of neutrinos  responsible
  %for the ultrahigh-energy cosmic rays,''
  Phys.\ Rev.\ Lett.\  {\bf 84}, 1378 (2000).
  %%CITATION = HEP-PH 9908276;%%

\bibitem{kt} For review, see 
  V.~A.~Kuzmin and I.~I.~Tkachev,
  %``Ultra high energy cosmic rays and inflation relics,''
  Phys.\ Rept.\  {\bf 320}, 199 (1999)
  [arXiv:hep-ph/9903542].
  %%CITATION = HEP-PH 9903542;%%

\bibitem{zg}
G.T.~Zatsepin, Dokl. Akad. Nauk. SSSR 80 (1951) 577; N.M.~Gerasimova and
G.T.~Zatsepin, Sov. Phys. JETP 11 (1960) 899.
  
\bibitem{kss}
Kusenko,Schissel,Stecker. unpublished

 
\bibitem{bahcall}
J.N.~Bahcall, Ann. Rev. Astron. Astrophys. 24 (1986) 577; 
J.A.~Frogel, \emph{ibid.}, 26 (1988) 51. 


 \bibitem{Kusenko:1997it}
  A.~Kusenko, M.~E.~Shaposhnikov, P.~G.~Tinyakov and I.~I.~Tkachev,
  %``Star wreck,''
  Phys.\ Lett.\ B {\bf 423}, 104 (1998).
%  [arXiv:hep-ph/9801212].
  %%CITATION = HEP-PH 9801212;%%


\end{thebibliography}
\end{document}